\documentclass[preprint,aps,amsmath,superscriptaddress,nofootinbib]{revtex4}
\usepackage{bm}
\usepackage{epsfig}

\newcommand{\nn}{\nonumber} 

\newcommand{\bea}{\begin{eqnarray}}
\newcommand{\eea}{\end{eqnarray}}

\newcommand{\Db}{\bar{D}}
\newcommand{\Hb}{\bar{H}}
\newcommand{\Pb}{\bar{P}}
\newcommand{\Vb}{\bar{V}}

\begin{document}



\title{Hadronic Decays of the $X(3872)$ to $\chi_{cJ}$ in Effective Field Theory} 

\author{Sean Fleming\footnote{Electronic address:     fleming@physics.arizona.edu}}
\affiliation{Department of Physics, 
        University of Arizona,
	Tucson, AZ 85721
	\vspace{0.2cm}}

\author{Thomas Mehen\footnote{Electronic address: mehen@phy.duke.edu}}
\affiliation{Department of Physics, 
	Duke University, Durham,  
	NC 27708\vspace{0.2cm}}

\date{\today\\ \vspace{1cm} }


\begin{abstract}

The decays of the $X(3872)$ to $P$-wave quarkonia are calculated under the assumption that it is a shallow bound state of
neutral charmed mesons. The $X(3872)$ is described  using an effective theory of nonrelativistic $D$ mesons and pions (X-EFT).
We calculate $X(3872)$ decays  by first matching heavy hadron chiral perturbation theory  (HH$\chi$PT) amplitudes for $D^0\,\bar{D}^{*0} \to \chi_{cJ} \,(\pi^0,\pi\pi)$  onto local operators in X-EFT, and then using these operators to calculate the 
$X(3872)$ decays. This procedure reproduces the factorization theorems  for $X(3872)$ decays to conventional quarkonia
previously derived using the operator product expansion. For single pion decays, we find  nontrivial dependence on the pion
energy from HH$\chi$PT diagrams with virtual $D$ mesons. This nontrivial energy dependence can potentially modify heavy quark
symmetry predictions for the relative sizes of decay rates. At leading order, decays to final states with two pions 
are dominated by the final state $\chi_{c1}\,\pi^0 \,\pi^0$, with a branching fraction just below that for the decay to $\chi_{c1}\, \pi^0$. 
Decays to all other final states with two  pions are highly suppressed.

\end{abstract}

\maketitle

\newpage

The $X(3872)$~\cite{Choi:2003ue,Acosta:2003zx,Abazov:2004kp} is a novel charmonium state that lies very close to the  $D^0 \bar{D}^{*0}$ 
threshold. The closeness of the $X(3872)$ to this threshold $(-0.6 \pm 0.6\, {\rm MeV})$ has prompted numerous 
authors to suggest that it is a $C=1$ molecular bound state of $D^0  \bar{D}^{*0} + \bar{D}^0 D^{*0}$,
for reviews see Refs.~\cite{Voloshin:2006wf,Godfrey:2008nc}.
The strongest evidence for the molecular hypothesis is the large ratio of the
 branching fractions for the $X(3872)$ decays to $J/\psi$ plus two or three pions~\cite{Abe:2005ix} 
\bea\label{isoval}
\frac{{\rm Br}[X(3872)\to J/\psi \,\pi^+ \pi^- \pi^0]}
     {{\rm Br}[X(3872)\to J/\psi \, \pi^+ \pi^-]}
= 1.0 \pm 0.4 \pm 0.3 \, .
\eea 
The final states have opposite G-parity which implies that the $X(3872)$ does not have definite isospin. Another possibility, suggested in 
Ref.~\cite{Gamermann:2007fi}, is that there are 
actually two nearly degenerate states
with opposite G-parity.
Measurements of the invariant mass distributions of the pions 
indicate that  these decays proceed through $J/\psi \,\rho$ for the 
$J/\psi \,\pi^+\, \pi^-$ final state, and through $J/\psi \, \omega$ for
the $J/\psi \,\pi^+ \,\pi^-\, \pi^0$ final state, suggesting that the $X(3872)$ couples to $I=0$ and $I=1$ channels with roughly equal strength. 
This is not possible for a conventional $c\bar{c}$ state. The observed branching ratio
can be understood if the $X(3872)$ is comprised primarily 
of neutral $D$ mesons, which is expected since the charged $D$-$D^*$ threshold
is far ($8.5 \,{\rm MeV}$) above the $X(3872)$ mass.

If the $X(3872)$ is a molecular state then the large distance behavior of the $D \bar{D}^*$
wavefunction is determined entirely by the binding energy and effective range theory (ERT)
can be used to calculate some properties of the $X(3872)$. ERT exploits 
the fact that at long distances the two-body wavefunction of the $X(3872)$ must be
\bea\label{molecularwfn}
\psi_{X}(r) \propto \frac{e^{-\gamma r}}{r} \, ,
\eea
where  $r$ is the separation of the meson constituents and  $\gamma =\sqrt{2 B \mu_{DD^*}}$,
where $B$ is the binding energy of the $X(3872)$, and $\mu_{DD^*}$ 
is $D^0$-$\bar{D}^{*0}$ reduced mass. Parameter-free predictions
for the decays $X(3872) \to D^0 \bar{D}^0 \pi^0$~\cite{Voloshin:2003nt} and 
$X(3872)\to D^0 \bar{D}^0 \gamma$~\cite{Voloshin:2005rt}
have been obtained using ERT. Experimental measurement of these decays consistent
with these calculations would confirm the molecular hypothesis. Unfortunately, data which 
can test these calculations is not currently available. For example, the photon
energy spectrum in the decay $X(3872) \to D^0 \bar{D}^0 \gamma$
near the kinematic endpoint can be computed in terms of the binding energy, but this spectrum
has not been measured yet. The ratio of branching ratios~\cite{Gokhroo:2006bt}
\bea
\frac{{\rm Br}[X(3872)  \to D^0 \,\bar{D}^0\, \pi^0]}{{\rm Br}[X(3872)  \to J/\psi \,\pi^+ \,\pi^-]} 
= 8.8^{+ 3.1}_{-3.6} \, ,
\eea
has been measured  but cannot be calculated  within ERT
since the decay $X(3872) \to J/\psi\, \pi^+ \, \pi^-$ is sensitive to the short-distance structure of the $X(3872)$.

Many of the observed decays of the $X(3872)$ are to final states that include conventional quarkonium such 
as $X(3872)\to J/\psi \,\pi^+\, \pi^-$, $X(3872)\to  J/\psi \,\pi^+\, \pi^-\, \pi^0$,
and $X(3872)\to J/\psi\, \gamma$. These decays necessarily involve shorter distance
scales where the description of the $X(3872)$ as a loosely bound state of $D$ and $\bar{D}^*$
mesons is no longer valid. To make quantitative calculations for these decays one can use factorization 
theorems that separate long-distance physics ($p \lesssim \gamma$) from physics at shorter 
distance scales ($p \gg \gamma$)~\cite{Braaten:2005jj,Braaten:2006sy}. 
Here $\gamma$ can be determined from the binding energy $B$ using the known masses
of the $D^0$, $D^{*0}$, and $X(3872)$. For  $m_{D^0}$ and $m_{D^{*0}}$, we use 
$1864.84 \pm 0.18 \, {\rm MeV}$ and $2006.97 \pm 0.19 \, {\rm MeV}$, respectively~\cite{Amsler:2008zz}.
For the $X(3872)$ we use the mass obtained in measurements of decays 
to $J/\psi \pi \pi$ final states only, which yields 
$m_X = 3871.2 \pm 0.5 \, {\rm MeV}$~\cite{Yao:2006px}.~\footnote{Interpretation of mass measurements in decays to $D^0 \bar{D}^0 \pi^0$ final states
is complicated by the possibility of two distinct states~\cite{Gamermann:2007fi,Amsler:2008zz} as well as threshold
enhancements due to the large scattering length in $D^0 \bar{D}^{*0}$ channel. We thank E. Braaten for discussion on this last point.}
Using these numbers we find $B = 0.6 \pm 0.6 \, {\rm MeV}$, so $\gamma \lesssim 48 \, {\rm MeV}$.
 The factorization theorems
express the rate for $X(3872)\to f$ (where $f$ is the final state) in terms of the the rate 
for $D^0 \bar{D}^{*0} \to f$ times universal factors that depend on the binding energy of the $X(3872)$. 
These factorization theorems have been used to analyze $X(3872)$ decays into $J/\psi$ plus light hadrons
in Ref.~\cite{Braaten:2005ai}. 

Recently, Ref.~\cite{Dubynskiy:2007tj} has studied the decays  $X(3872) \to \chi_{cJ} (\pi^0, \pi \pi)$
 and argued that measurement of the partial rates for these decays can discriminate between
conventional charmonium and molecular  interpretations of the $X(3872)$. Decays to $\chi_{cJ}$ are particularly interesting 
since the $\chi_{cJ}$ are in a heavy quark multiplet and therefore heavy quark symmetry
can be used to make predictions for the relative sizes of the partial rates.

In this paper, we study the strong decays $X(3872) \to \chi_{cJ} (\pi^0,\pi \pi)$ using a combination of heavy hadron chiral 
perturbation theory (HH$\chi$PT)~\cite{Wise:1992hn,Burdman:1992gh,Yan:1992gz} and X-EFT, an effective field theory for the
$X(3872)$ developed in Ref.~\cite{Fleming:2007rp}. HH$\chi$PT is an effective theory for low energy hadronic interactions of
mesons containing heavy quarks that incorporates the constraints of the heavy quark and chiral symmetries of QCD. We use
HH$\chi$PT to calculate the $D^0 \bar{D}^{*0} \to \chi_{cJ}(\pi^0, \pi \pi)$\footnote{Here and below decays related by charge
conjugation are implied.} transition amplitudes. 
X-EFT~\cite{Fleming:2007rp} generalizes the effective theory developed for shallow nuclear 
bound states in Ref.~\cite{Kaplan:1998tg,Kaplan:1998we,vanKolck:1998bw}. In this theory, modes have momenta of order $\gamma$
so the degrees of freedom are non-relativistic charmed mesons and non-relativistic pions. The decays 
$X(3872) \to \chi_{cJ} (\pi^0,\pi \pi)$ proceed through local  operators that couple the $D^0$ and $\bar{D}^{*0}$ to the 
 $\chi_{cJ}$ and one or two pions. We use the HH$\chi$PT calculation of the $D^0 \bar{D}^{*0} \to \chi_{cJ}(\pi^0, \pi \pi)$ 
amplitudes to fix the coefficients of these operators in the X-EFT Lagrangian, then calculate the decays of the $X(3872)$. 
The resulting decay rates are consistent with the factorization theorems for 
$X(3872)$ decays proven in Refs.~\cite{Braaten:2005jj,Braaten:2006sy}.

In the leading order (LO) diagrams for $D^0 \bar{D}^{*0} \to \chi_{cJ}\,
\pi^0$, virtual charmed meson propagators introduce dependence on the pion momentum which modifies the usual $p_\pi^3$
dependence expected for a $P$-wave single pion decay. This can give significant  corrections to the relative rates
predicted in Ref.~\cite{Dubynskiy:2007tj}. We also calculate the differential decay rates for $X(3872) \to \chi_{cJ} \, \pi \,\pi$. At LO in
HH$\chi$PT, amplitudes for $D^0 \bar{D}^{*0} \to \chi_{cJ}\, \pi^0$ and $D^0 \bar{D}^{*0} \to \chi_{cJ} \, \pi \, \pi$ are
given in terms of a single unknown coupling in the  HH$\chi$PT lagrangian. Therefore, at  LO, the  normalization of the
$X(3872) \to \chi_{cJ} \, \pi \, \pi $ decay rates  can be given in terms of  the rates for  $X(3872) \to \chi_{cJ} \, \pi^0$.
We find that $X(3872) \to \chi_{c1}\, \pi^0 \, \pi^0$ is the dominant decay mode with two pions in the final state, with a branching fraction that is
just $0.6$  times the branching fraction for $X(3872) \to \chi_{c1} \, \pi^0$.
All other decays to final states with two pions are considerably smaller. In some of the tree level diagrams for
$D^0 D^{*0} \to \chi_{c1} \pi^0 \pi^0$, a virtual $D^0$ meson can go nearly on-shell, with the potential divergence cutoff 
by the width of the $D^{*0}$. The resulting contribution for the decay rate is enhanced by $\sim 2\pi E_\pi/\Gamma_{D^{*0}} \sim 10^4$,  
where $E_\pi$ is the typical energy of the pion in the final state (about 200 MeV) and $\Gamma_{D^{*0}}$, the width of the 
$D^{*0}$, is estimated to be about 
68 keV~\cite{Hu:2005gf}. This enhancement only occurs for the $\chi_{c1} \pi^0 \pi^0 $ final state, total partial decay rates
to other final states with two pions are smaller by orders of magnitude.

The LO HH$\chi$PT lagrangian for mesons containing heavy quarks or antiquarks at rest is 
\bea
{\cal L} &=& {\rm Tr}[H^\dagger_a (i D_0)_{ba} H_b] - g\, {\rm Tr}[H^\dagger_a  H_b \,  \vec{\sigma} \cdot  \vec{A}_{ba}]
+\frac{\Delta_H}{4}{\rm Tr}[H^\dagger_a \,  \sigma^i \, H_a \, \sigma^i] \,  \nn \\
&+&{\rm Tr}[\Hb^\dagger_a (i D_0)_{ab} \Hb_b] + g \,{\rm Tr}[\Hb^\dagger_a   \, \vec{\sigma}\cdot \vec{A}_{ab} \Hb_b]
+\frac{\Delta_H}{4}{\rm Tr}[\Hb^\dagger_a \,  \sigma^i \, \Hb_a \, \sigma^i] \, .
\eea
We use the two component notation of Ref.~\cite{Hu:2005gf}. The field $H_a$ is given by
\bea
H_a =\vec{V}_a \cdot {\vec{\sigma}}+ P_a \, ,
\eea
where $\vec{V}_a$ annihilates $D^*_a$ mesons and $P_a$ annihilates $D_a$ mesons. The subscript
$a$ is an $SU(3)$ index, and $a=1$ for neutral $D$ mesons. The field for antimesons is
\bea
\Hb_a = -{\vec{\Vb}}_a \cdot {\vec{\sigma}} + \Pb_a \, .
\eea
The 
field ${\vec A}_{ab}$ is the axial current of chiral perturbation theory, $\vec{A}_{ab} = - \vec{\nabla} \pi_{ab}/f_\pi + ...$, 
where $f_\pi$ is the pion decay constant and $\pi_{ab}$ are the Goldstone boson fields.
The axial coupling, $g$, is taken to be $0.6$ and $\Delta_H$ is the  hyperfine splitting. 

To calculate the amplitude for $D^0 \, \bar{D}^{*0} \to \chi_{cJ} \,\pi^0$
we need to include the $\chi_{cJ}$ in the HH$\chi$PT lagrangian. 
The $\chi_{cJ}$ are degenerate in the heavy quark limit and form a heavy quark multiplet.
The fields for $\chi_{cJ}$ have been constructed in a covariant formalism in 
Ref.~\cite{Casalbuoni:1992dx}. In our two component notation, the $\chi_{cJ}$ fields 
are represented by 
\bea
\chi^i &=& \sigma^j \, \chi^{ij} \nn \\
&=& \sigma^j\left(\chi^{ij}_2 + \frac{1}{\sqrt{2}}\,\epsilon^{ijk} \chi^k_1 + \frac{\delta^{ij}}{\sqrt{3}} \chi_0 \right)
\, .
\eea
The transformation properties of the heavy quark and quarkonium fields under the various symmetries are:
\bea
\begin{array}{rccc}
{\rm rotations:} & \qquad H_a \to U H_a  U^\dagger & \qquad \Hb_a \to U\Hb_a  U^\dagger & \qquad \chi^i \to R^{ij} U^\dagger \chi^j U \nn \\
{\rm heavy \,quark \,spin:} &\qquad H_a \to S H_a &\qquad \Hb_a \to \Hb_a \bar{S}^{\dagger} 
&\qquad \chi^i \to S \chi^i \bar{S}^\dagger  \nn \\
{\rm parity:}& \qquad  H_a \to  - H_a  &\qquad  \Hb_a \to  - \Hb_a  &\qquad \chi^i \to \chi^i\nn \\
{\rm charge\, conjugation:}&
\qquad  H_a \to  \sigma_2 \Hb_a^{\, T} \sigma_2  &\qquad  \Hb_a \to  \sigma_2 H_a^{\,T} \sigma_2 &\qquad \chi^i \to -\sigma_2 (\chi^i)^T \sigma_2 = \chi^i\nn \nn \\
SU_L(3)\times SU_R(3): &\qquad  H_a \to  H_b V_{ba}^\dagger 
&\qquad  H_a \to  V_{ab} \Hb_b & \qquad \chi^i \to \chi^i 
\end{array}
\eea
Note that under charge conjugation, the components of $H_a$ transform as $\vec{{V}}_a \to {\vec{\Vb}}_a$ and $P_a \to \Pb_a$. 
Therefore, the $C$ even combination of $D$ mesons that couples to the $X(3872)$ is ${\vec{V}}_1 \Pb_1 + {\vec{\Vb}}_1 P$.
In the first line, $U$($R^{ij}$) are $2\times 2$ ($3\times 3$) rotation matrices related by
 $U^\dagger \sigma^i U = R^{ij} \sigma^j$. In the second line, $S$ is a rotation matrix acting only 
on the heavy-quark spin, $\bar{S}$ is a rotation matrix acting only 
on the heavy-antiquark spin  and in the last line $V_{ab}$ is an $SU(3)$ matrix.

\begin{figure}[t]
 \begin{center}
 \includegraphics[width=5.0in]{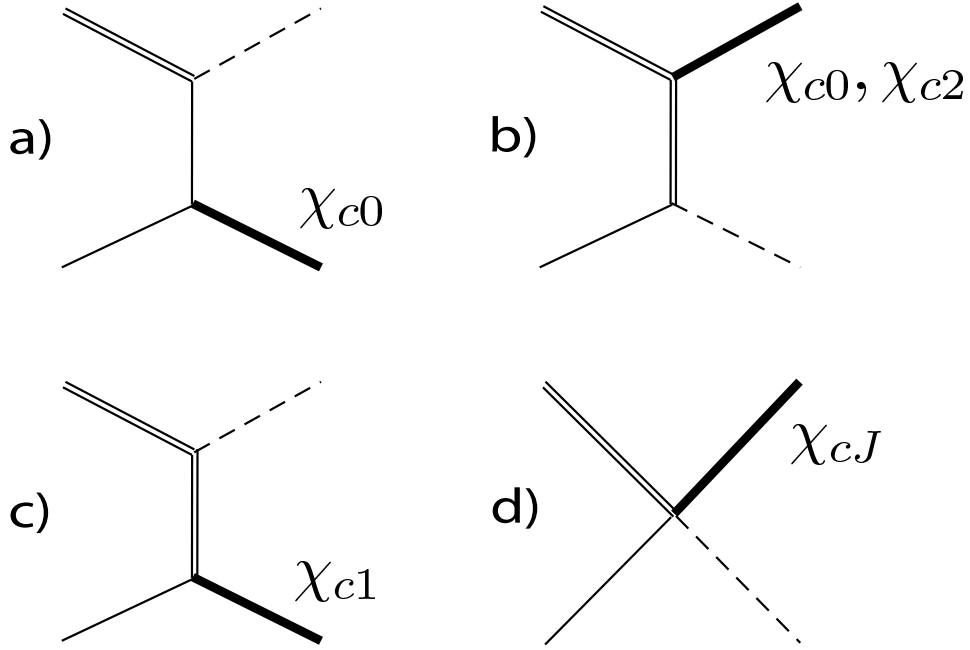}
 \end{center}
 \vskip -0.7cm \caption{Feynman diagrams contributing to the 
$D^0 \bar{D}^{*0} \to \chi_{cJ} \, \pi^0$ 
amplitude. The solid line is a $D^0$ meson, the
double line is a $\bar{D}^{*0}$ meson, the dotted line is a $\pi^0$ meson, and the thick solid line
is the $\chi_c$ meson.}
  \label{ddchipi}
 \end{figure}

The lagrangian coupling the $\chi_{cJ}$ to heavy mesons is
\bea\label{lagchi}
{\cal L}_\chi = i\frac{g_1}{2} {\rm Tr}[\chi^{\dagger \, i}  H_a \sigma^i \, \Hb_a] 
+ \frac{c_1}{2} {\rm Tr}[ \chi^{\dagger\, i} H_a \sigma^j \,\Hb_b] \epsilon_{ijk} A^k_{ab}+{\rm h.c.} \, .
\eea
The leading contribution to the amplitude for $D^0 \, \bar{D}^{*0}  \to \chi_{cJ}\, \pi^0$ comes from 
the tree level diagrams in Figs.~\ref{ddchipi}a), b), and c). In these diagrams, the coupling of $\chi_{cJ}$ to $D$ 
mesons comes from the first operator in Eq.~(\ref{lagchi}). Fig.~\ref{ddchipi}a) contributes to $\chi_{c0}$ decays, 
Fig.~\ref{ddchipi}b) contributes to both $\chi_{c0}$ and $\chi_{c2}$ decays, and Fig.~\ref{ddchipi}c) contributes to 
$\chi_{c1}$ decays only. In the HH$\chi$PT power counting, one takes $E_\pi \sim m_\pi \sim \Delta_H \sim Q$, and expands 
amplitudes in powers of $Q$. The graphs in 
Figs.~\ref{ddchipi}a)-c) are $O(Q^0)$ since the coupling of the $D$ mesons to $\chi_{cJ}$  has no derivatives and is therefore $O(Q^0)$,  the axial coupling of the $D$ mesons is $O(Q)$, and the heavy meson propagator is $O(Q^{-1})$.  Fig.~\ref{ddchipi}d) shows the contribution to the amplitude from the second operator in Eq.~(\ref{lagchi}). This contribution
is $O(Q)$ because this operator contains one derivative. Chiral loop corrections to the LO diagrams in Eq.~(\ref{ddchipi}) are suppressed by $Q^2$ and therefore subleading to the diagram
in Fig.~\ref{ddchipi}d). It is also clear that higher dimension operators that couple the $D$ mesons to $\chi_{cJ}$ 
either contain two spatial derivatives or an insertion of the light quark mass matrix and are suppressed relative to the 
leading operator in Eq.~(\ref{lagchi}) by at least $Q^2$. Therefore the diagrams in Fig.~\ref{ddchipi} give the complete 
HH$\chi$PT calculation to next-to-leading (NLO) order.
  
We obtain 
\bea\label{amp}
{\cal M}(D^0 \Db^{*0} \to \chi_{c0} \pi^0) &=& 
\frac{2 \,\vec{\epsilon}_D\cdot \vec{p}_\pi}{\sqrt{3} \,f_\pi} \, \left[\sqrt{2} \,g \, g_1  
\left(\frac{3}{4} \frac{1}{E_\pi -\Delta_H} +\frac{1}{4}\frac{1}{E_\pi +\Delta_H}\right) 
+  \frac{c_1}{\sqrt{2}}  \right]  \, ,\nn \\
{\cal M}(D^0 \Db^{*0}  \to \chi_{c1} \pi^0) &=& 
\frac{\vec{\epsilon}^{\,\,*}_1\cdot (\vec{p}_\pi \times\vec{ \epsilon}_D)}{\sqrt{2} \, f_\pi} \,\left[ \sqrt{2}\,g \, g_1  
\frac{1}{E_\pi} + \frac{c_1}{\sqrt{2}}  \right] \, ,\nn \\
{\cal M}(D^0 \Db^{*0}\to \chi_{c2} \pi^0) &=& 
\frac{-(\epsilon_2^*)^{ij} (p_\pi)_i (\epsilon_D)_j}{f_\pi} \, \left[  \sqrt{2} \,g \, g_1 
\frac{1}{E_\pi+\Delta_H} + \frac{c_1}{\sqrt{2}}  \right]  \, .
\eea
The results for the charge conjugate initial states are identical.
These amplitudes must be multiplied by a factor of $\sqrt{ m_D \, m_{D^*} \, m_{\chi_{cJ}}}$ to account for
the non-relativistic normalization of the HH$\chi$PT fields.

The virtual charmed mesons in Figs.~\ref{ddchipi}a)-c) are off-shell by $\sim E_\pi$,
where $E_\pi$ varies from 432 MeV to 305 MeV in the three decays. X-EFT contains
non-relativistic $D^0$, $D^{*0}$, and $\pi^0$, and total energies of these particles are assumed to be within a few MeV of
the $D^0 \bar{D}^{*0}$ threshold. Thus we must match the tree level diagrams in Fig.~\ref{ddchipi} onto local operators  in
X-EFT. We will illustrate this for the decay to $\chi_{c0}$, the generalization to other decays is
straightforward. The operator  in X-EFT which reproduces the amplitudes for $D^{0}\, \bar{D}^{*0} \to  \chi_{c0} \,\pi^0$
and $\bar{D}^{0} \,D^{*0} \to \chi_{c0}\, \pi^0$ is
\bea\label{Lagdecay}
{\cal L} = i \frac{C_{\chi, 0}(E_\pi)}{4\sqrt{m_\pi}} \, \frac{2}{\sqrt{3}} \,
(V^i \, \Pb +\Vb^i \, P) \, \frac{ \nabla_i \pi^0}{f_\pi} \, \chi_{c0}^\dagger \, ,
\eea
where we have dropped the $SU(3)$ labels on the $D$ meson fields that are  implicitly neutral. 
Note, when evaluating the amplitude we must multiply the naive Feynman rule obtained 
from Eq.~(\ref{Lagdecay}) by $\sqrt{16\,m_\pi\, m_{\chi_{c0}}\, m_D \, m_{D^*}}$
to account for the normalizations
of the fields in the effective theory, which are all nonrelativistic, including the pion. The factor
of $1/(4\sqrt{m_\pi})$ is included to compensate for the differing normalizations in the effective field
theory and HH$\chi$PT.
Comparing with
Eq.~(\ref{amp}) (times $\sqrt{m_D \, m_{D*} \,m_{\chi_{c0}}}$), we find for $C_{\chi,0}(E_\pi)$,
\bea
C_{\chi,0}(E_\pi) = \sqrt{2} \,g \, g_1 \,
\left(\frac{3}{4} \frac{1}{E_\pi -\Delta_H} +\frac{1}{4}\frac{1}{E_\pi +\Delta_H}\right) 
+ \frac{c_1}{\sqrt{2}}   \, .
\eea

\begin{figure}[t]
 \begin{center}
 \includegraphics[width=2.5in]{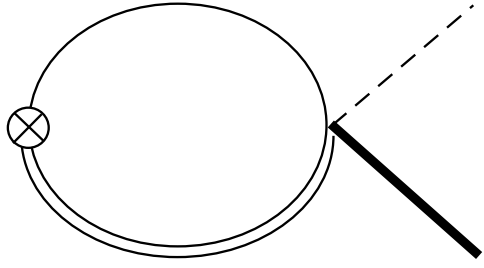}
 \end{center}
 \vskip -0.7cm \caption{Feynman diagram contributing to the 
$X(3872) \to \chi_{cJ} \,\pi^0$ decay amplitude. The circle with cross represents the interpolating 
field for the $X(3872)$.  }
\label{Xchipi}
 \end{figure}
Now we consider the decay $X(3872) \to \chi_{cJ}\, \pi^0$ mediated by this operator. The one-loop diagram
contributing to this decay is depicted in Fig.~\ref{Xchipi}.
The decay matrix element is obtained from the diagram in Fig.~\ref{Xchipi}  after dividing by the wavefunction 
renormalization factor $\mu_{DD^*}/\sqrt{2 \pi \gamma}$~\cite{Fleming:2007rp}. After summing over both channels, 
the result is 
\bea
{\cal M}(X(3872)\to \chi_{c0}\pi^0) = -\frac{1}{C_0(\Lambda_{PDS})}\, \frac{\sqrt{2 \pi \gamma}}{\mu_{DD^*}}\,
\sqrt{m_{\chi 0}\, m_X} \, C_{\chi, 0}(E_\pi) \, \frac{2}{\sqrt{3}}\,\vec{ p}_\pi \cdot \vec{\epsilon}_X \, .
\eea
In evaluating the diagram in Fig.~\ref{Xchipi} we have used
\bea
2 \mu_{DD^*} \int \frac{d^D l}{(2\pi)^D} \frac{1}{l^2 + \gamma^2} 
&=& \frac{\mu_{DD^*}}{2\pi} \, (\Lambda_{\rm PDS}- \gamma) \nn \\
&=&-\frac{1}{C_0(\Lambda_{\rm PDS})} \, .
\eea

It is straightforward to evaluate the rates:
\bea\label{decayrates}
\Gamma[X(3872)\to \chi_{cJ}\pi^0] = \frac{2 \pi \gamma}{\mu^2_{DD^*} C^2_0(\Lambda_{\rm PDS})} \, 
\frac{m_{\chi_{cJ}}}{m_X}\, \frac{p_\pi^3}{72 \pi f_\pi^2}\, \alpha_J \, |C_{\chi,J}(E_\pi)|^2 \, ,
\eea
where $\alpha_0 =4$, $\alpha_1 = 3$, $\alpha_2=5$, and the functions $C_{\chi,J}(E_\pi)$ are given by
\bea
C_{\chi,0}(E_\pi) &=& \sqrt{2} \,g \, g_1
\left(\frac{3}{4} \frac{1}{E_\pi -\Delta_H} +\frac{1}{4}\frac{1}{E_\pi +\Delta_H}\right) 
+ \frac{c_1}{\sqrt{2}}    \,  ,\nn \\
C_{\chi,1}(E_\pi)  &=& \sqrt{2} \,g \, g_1 \frac{1}{E_\pi} + \frac{c_1}{\sqrt{2}} \, ,\nn \\
C_{\chi,2}(E_\pi)  &=& \sqrt{2} \,g \, g_1 \frac{1}{E_\pi+\Delta_H} + \frac{c_1}{\sqrt{2}} \, .
\eea

Using these results we can calculate the partial rates for $\Gamma[X(3872) \to \chi_{cJ}\,\pi^0]$. At LO
($c_1=0$) we find (denoting $\Gamma[X(3872) \to \chi_{cJ} \, \pi^0] \equiv \Gamma_J$)
\bea\label{lopred}
\Gamma_0:\Gamma_1:\Gamma_2::4.76:1.57:1 \, .
\eea
The energy dependence from the virtual pion propagators makes a significant change in the predictions
for the relative sizes of the partial rates. The NLO predictions depend on the ratio $c_1/g_1$ which is  undetermined.
The ratio $c_1/g_1$ has dimensions of inverse mass, when $c_1/g_1$ is varied from 
$(100 \,{\rm MeV})^{-1}$ to $(1000 \,{\rm MeV})^{-1}$, we find:
\bea\label{nlopred}
\Gamma_0:\Gamma_1:\Gamma_2&::&3.01:1.06:1 \qquad c_1/g_1=  (100 \,{\rm MeV})^{-1} \nn \\
\Gamma_0:\Gamma_1:\Gamma_2&::&3.49:1.20:1 \qquad c_1/g_1=  (300 \,{\rm MeV})^{-1} \nn \\
\Gamma_0:\Gamma_1:\Gamma_2&::&3.76:1.28:1 \qquad c_1/g_1=  (500 \,{\rm MeV})^{-1} \nn \\
\Gamma_0:\Gamma_1:\Gamma_2&::&4.11:1.38:1 \qquad c_1/g_1=  (1000 \,{\rm MeV})^{-1} \, .
\eea
For the largest values of $c_1/g_1$, we obtain predictions for the relative sizes of the partial rates
similar to those found in Ref.~\cite{Dubynskiy:2007tj}. As $c_1/g_1$ decreases, the predictions tend to those appearing
in Eq.~(\ref{lopred}). Experimental measurement of the relative sizes of the partial rates 
for decays to $\chi_{c,J}\,\pi^0$ can be used to determine $c_1/g_1$.

\begin{figure}[t]
 \begin{center}
 \includegraphics[width=4.5in]{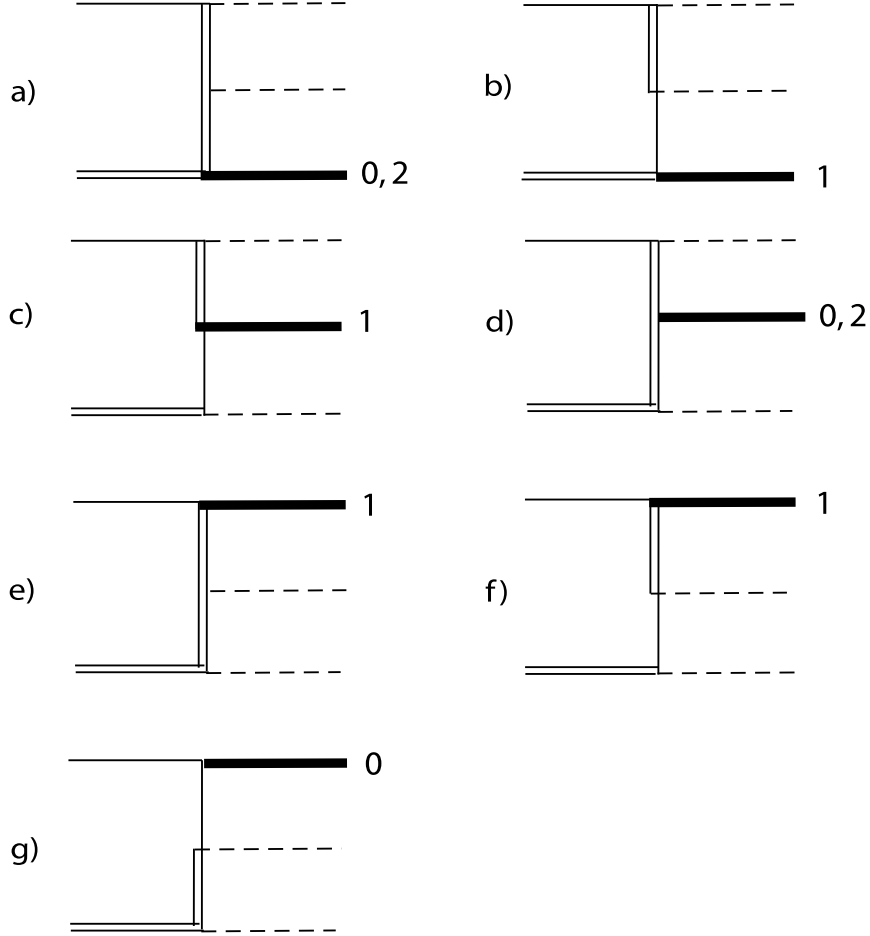}
 \end{center}
 \vskip -0.7cm \caption{Feynman diagrams contributing to the 
$D^0 \bar{D}^{*0} \to \chi_{cJ} \pi^0 \pi^0$ amplitude. Diagrams with the final state pions
crossed are not shown. The numbers denote the angular momentum quantum number of the $\chi_{cJ}$
produced in the decay.}
\label{Xchipipi}
 \end{figure}

Refs.~\cite{Braaten:2005jj,Braaten:2006sy} derived factorization theorems for $X(3872)$ decaying into 
quarkonium states and light hadrons. Note that the first term in the decay rates in Eq.~(\ref{decayrates})
can be written as
\bea\label{ldme}
\frac{2 \pi \gamma}{\mu_{DD^*}^2 C_0(\Lambda_{\rm PDS})^2} 
= \frac{1}{3}\sum_\lambda |\langle 0| \frac{1}{\sqrt{2}}{\epsilon}_i(\lambda) 
\,(V^i \, \Pb +\Vb^i \, P) |X(3872,\lambda)\rangle|^2 \, ,
\eea 
where $\vec{\epsilon}(\lambda)$ is a polarization vector. Thus we see that the expressions for the
partial widths factor into the product of long-distance matrix elements, given in Eq.~(\ref{ldme}),
times short-distance coefficients, which are proportional to the cross sections for $D^0\,\bar{D}^{*0} \to 
\chi_{cJ}\, \pi^0$. This is the content of the factorization theorems of
Refs.~\cite{Braaten:2005jj,Braaten:2006sy}. A main point of this paper is that HH$\chi$PT 
can be used to calculate the short-distance coefficients in these factorization theorems.

Next we turn to the calculation of the three-body decays, $X(3872) \to \chi_{cJ}\, \pi^0 \,\pi^0$. 
In Fig.~\ref{Xchipipi} we
show the diagrams contributing to the transition amplitudes for $D^0 \, \bar{D}^{*0} \to \chi_{cJ}\, \pi^0 \, \pi^0$. These amplitudes are matched onto
local operators in X-EFT, then these local operators are used to calculate the decays  $X(3872)\to \chi_{cJ} \, \pi^0
\, \pi^0$, in a manner identical to the procedure used to calculate  $X(3872) \to \chi_{cJ} \, \pi^0$. Here we give our
results for the differential partial decay rates. 

The differential decay rate for $X(3872) \to \chi_{c 0} \pi^0 \pi^0$ is given by
\bea\label{dr1}
\frac{d^2 \Gamma[X \to \chi_{c 0} \, \pi^0 \, \pi^0]}{dE_1 dE_2} 
= \frac{2 \pi \gamma}{\mu_{DD^*}^2 C_0(\Lambda_{\rm PDS})^2}
\frac{\pi}{18}\frac{g^4 g_1^2 m_{\chi_{c0}}}{\Lambda_\chi^4}
\left( p_1^2 \, p_2^2 - (\vec{p}_1 \cdot \vec{p}_2)^2 \right) 
\,F^2(E_1,E_2)\,,
\eea
where
\bea
F(E_1,E_2) = (E_2-E_1)\!\!\!\!
&&\left[\frac{1}{(E_1+\Delta_H)(E_2+\Delta_H)(E_1+E_2+\Delta_H)}  \right.\nn \\
&&\left.
+\frac{3}{E_1 \, E_2 \,(E_1+E_2-\Delta_H)} 
+\frac{\Delta_H}{E_1 \, E_2\, (E_1+\Delta_H)(E_2+\Delta_H)}\right] \, .
 \eea
Here, $E_i (\vec{p}_i)$ refers to the energy (three-momentum) of one of the $\pi^0$ and $p_i 
= |\vec{p}_i|$. The partial decay rates are symmetric under $1 \leftrightarrow 2$. In these decays
$\Delta_H$ is equal to the neutral hyperfine splitting, 
$\Delta_H= \Delta_H^{00} \equiv m_{D^{*0}}-m_{D^0} = 142 \, {\rm MeV}$. 
 
The rates for $X(3872) \to \chi_{cJ}\, \pi^+ \pi^-$ can also 
be calculated. The differential rates for $X(3872) \to \chi_{cJ} \,\pi^+ \,\pi^-$ are bigger by a 
factor of two than the corresponding rates for $X(3872) \to \chi_{cJ} \,\pi^0 \,\pi^0$ due to 
isospin factors in the diagram. Furthermore,  the factor of $\Delta_H$ can be $\Delta_H^{00}, 
\Delta_H^{0+}\equiv m_{D^{*0}}-m_{D^+} = 137 \, {\rm MeV}$, or $\Delta_H^{+0} \equiv m_{D^{*+}}-m_{D^0} = 145 \, {\rm MeV}$,
depending on the diagram, so the expressions above (times 2) can only be used if one neglects isospin breaking in the heavy meson
hyperfine splittings.

The differential decay rate for $X(3872) \to \chi_{c 1} \pi^0 \pi^0$ obtained from diagrams in Figs.~\ref{Xchipipi}b),c),e), and f)  is
\bea\label{chi1ddr}
\frac{d^2 \Gamma[X \to \chi_{c 1}\,  \pi^0 \, \pi^0]}{dE_1 dE_2} 
&=& \frac{2 \pi \gamma}{\mu_{DD^*}^2 C_0(\Lambda_{\rm PDS})^2}
\frac{\pi}{3}\frac{g^4 g_1^2 m_{\chi_{c1}}}{\Lambda_\chi^4}  \\
&&\hspace{-1cm} \times
\bigg[p_1^2 \, p_2^2 \,(B_{12}^2 +B_{21}^2) +
(\vec{p}_1 \cdot \vec{p}_2)^2 \left(3 \,A^2 
+ 2 \,A\,(B_{12} +B_{21}) +2\, B_{12} \, B_{21} \right)  
\bigg]\, , \nn
\eea
where
\bea
A &\equiv& A(E_1,E_2) = \frac{1}{E_1 \, E_2} +
\frac{E_1+E_2+2 \Delta_H}{(E_1 + E_2)(E_1+\Delta_H)(E_2+\Delta_H)} \, , \nn \\
B_{12} &\equiv& B(E_1,E_2) = \frac{2}{(E_1+\Delta_H)(E_2 -\Delta_H)} +
\frac{2 E_1+\Delta_H}{E_1 \,(E_1+E_2) (E_1+\Delta_H) }\, , \nn \\
B_{21} &\equiv& B(E_2,E_1)\,.
 \eea
In Figs.~\ref{Xchipipi}c) and \ref{Xchipipi}f), the virtual $D^0$ meson can go on-shell leading to the poles of $E_i -\Delta_H$ 
in the expression for the decay rate. In the case of $\pi^+ \, \pi^-$ final states, the virtual charged $D$ meson cannot go on-shell 
but $E_i-\Delta_H$ can be as small as 2 MeV. The pole in the decay amplitude is regulated by putting the 
external $D^{*0}$ on its complex energy mass shell, $E_{D^{*0}}=\Delta_H - i \Gamma_{D^{*0}}/2$.~\footnote{We thank
Eric Braaten for suggesting this.} The width of the $D^{*0}$ has not been measured. If one uses HH$\chi$PT
to relate the width of the $D^{*0}$ to the measured width of the $D^{*+}$ and the measured strong decay branching 
fractions for both $D^{*0}$ and $D^{*+}$, one finds $\Gamma_{D^{*0}} =0.068 \,{\rm MeV}$~\cite{Hu:2005gf}. Since 
this scale is orders of magnitude smaller than all other scales in the problem, we will keep only terms 
which do not vanish in the limit $\Gamma_{D^{*0}}\to 0$. For terms with single poles in $E_i - \Delta_H$,
this amounts to using a principal value prescription. When integrating over the double poles, 
we replace $(E_i-\Delta_H)^2 \to (E_i - \Delta_H)^2 +\Gamma_{D^{*0}}^2/4$ in the denominators of Eq.~(\ref{chi1ddr}),
then perform the energy integrals using the formula
\bea\label{prescrip}
\int_{x_0}^{x_1} dx \,\frac{f(x)}{x^2 +\Gamma^2/4} &=& \frac{2\pi}{\Gamma}\, f(0)\,\theta(-x_0)\theta(x_1)
+\left(-\frac{1}{x_1} + \frac{1}{x_0}\right) f(0) \nn \\
&&+ f^\prime(0)\frac{1}{2} \log\left(\frac{x_1^2}{x_0^2}\right)
+ \int_{x_0}^{x_1} dx\,\frac{f(x)- f(0) -x \,f^\prime(0)}{x^2} + ... \, ,
\eea
where the elipsis denotes terms that can be neglected when $\Gamma \ll x_0,x_1$. 
Keeping only the first term in Eq.~(\ref{prescrip}) is called the narrow width approximation
which turns out to be an excellent approximation for $X(3872) \to \chi_{c1} \pi^0 \pi^0$.

One might also worry about contributions from interference terms that contain single poles in $E_1$ and 
$E_2$. These are of the form 
\bea
&&{\rm Re}\left(\frac{1}{E_1-\Delta_H - i \Gamma/2} \frac{1}{E_2-\Delta_H + i \Gamma/2}\right) \\
&=& \frac{(E_1-\Delta_H)(E_2-\Delta_H)}{( (E_1-\Delta_H)^2 + \Gamma^2/4)((E_2-\Delta_H) + \Gamma^2/4)} 
 +\frac{\Gamma^2/4}{( (E_1-\Delta_H)^2 + \Gamma^2/4)((E_2-\Delta_H) + \Gamma^2/4)} \nn \,.
\eea
The first term on the righthand side clearly yields two  single poles which should be dealt with via the principal value 
prescription in the limit $\Gamma_{D^{*0}} \to 0$. The second term only has a nonvanishing contribution (in the limit 
$\Gamma_{D^{*0}} \to 0$) with support at $E_1 = E_2 = \Delta_H$, which lies outside available phase space.

Finally, the differential decay rate for $X(3872) \to \chi_{c 2} \pi^0 \pi^0$ is 
\bea
\frac{d^2 \Gamma[X \to \chi_{c 2} \, \pi^0 \, \pi^0]}{dE_1 dE_2} 
&=& \frac{2 \pi \gamma}{\mu_{DD^*}^2 C_0(\Lambda_{\rm PDS})^2}
\frac{2\pi}{3}\frac{g^4 g_1^2 m_{\chi_{c2}}}{\Lambda_\chi^4}  \\
&&\hspace{-1cm} \times
\bigg[p_1^2 \, p_2^2 \,\left(\frac{5}{3}\, \left(C^2 + C\,(D_{12}-D_{21})\right)
+ \frac{7}{6} (D_{12}^2 + D_{21}^2) +\frac{2}{3}\, D_{12}\,D_{21}\right)
\, , \nn \\
&&\hspace{-1cm} +
(\vec{p}_1 \cdot \vec{p}_2)^2 \left( -\frac{5}{3}\, \left(C^2 + C\,(D_{12}-D_{21})\right)
- \frac{1}{6} (D_{12}^2 + D_{21}^2) +\frac{4}{3}\, D_{12}\,D_{21}\right) \bigg] \, . \nn
\eea
where
\bea\label{dr6}
C&\equiv& C(E_1,E_2) = \frac{E_2-E_1}{(E_1+\Delta_H)(E_2+\Delta_H)(E_1 + E_2+\Delta_H)} \, , \nn \\
D_{12} &\equiv& D(E_1,E_2) = \frac{1}{(E_1+\Delta_H)E_2}  \, , \nn \\
D_{21} &\equiv& D(E_2,E_1)\,.
 \eea 

The normalization of the 
three-body decay rates is unknown but can be given in terms of the LO 
chiral perturbation theory expressions for the two-body decay rates computed earlier.
Integrating over three-body phase space we find 
\bea\label{ratios}
\left(\frac{{\rm Br}[X(3872) \to \chi_{c 0} \, \pi^0 \, \pi^0]}{{\rm Br}[X(3872) \to \chi_{c 0} \, \pi^0 ]}\right)_{\rm LO}
 = 9.1 \, 10^{-6} \, ,\nn \\
\left(\frac{{\rm Br}[X(3872) \to \chi_{c 1} \, \pi^0 \, \pi^0]}{{\rm Br}[X(3872) \to \chi_{c 1} \, \pi^0 ]}\right)_{\rm LO}
 = 6.1\, 10^{-1} \, , \nn \\
\left(\frac{{\rm Br}[X(3872) \to \chi_{c 2} \, \pi^0 \, \pi^0]}{{\rm Br}[X(3872) \to \chi_{c 2} \, \pi^0 ]}\right)_{\rm LO}
 = 7.8\, 10^{-6} \, .
 \eea
 Here we have used $m_X= 3871.4 \, {\rm MeV}$ and $\Lambda_\chi = 4 \pi f_\pi = 1659 \, {\rm MeV}$.
The subscript LO emphasizes that this result only holds at LO in $\chi$PT. At NLO the prediction
depends on the unknown parameter, $c_1$. 

We have not numerically computed the corresponding branching fraction ratios for $X(3872) \to \chi_{cJ} \pi^+ \pi^-$.
For final states with $J=0$ or $2$, isospin violation is a small effect so we expect
\bea
\left(\frac{{\rm Br}[X(3872) \to \chi_{c 0} \, \pi^+ \, \pi^-]}{{\rm Br}[X(3872) \to \chi_{c 0} \, \pi^0 ]}\right)_{\rm LO}
\approx 2 \left(\frac{{\rm Br}[X(3872) \to \chi_{c 0} \, \pi^0 \, \pi^0]}{{\rm Br}[X(3872) \to \chi_{c 0} \, \pi^0 ]}\right)_{\rm LO}
\approx 1.8 \,10^{-5} \nn \, , \\
 \left(\frac{{\rm Br}[X(3872) \to \chi_{c 2} \, \pi^+ \, \pi^-]}{{\rm Br}[X(3872) \to \chi_{c 2} \, \pi^0 ]}\right)_{\rm LO}
\approx 2 \left(\frac{{\rm Br}[X(3872) \to \chi_{c 2} \, \pi^0 \, \pi^0]}{{\rm Br}[X(3872) \to \chi_{c 2} \, \pi^0 ]}\right)_{\rm LO}
\approx 1.6 \,10^{-5}\, .
\eea
For final states with $\chi_{c1} \, \pi^+ \, \pi^-$, isospin breaking effects are crucial since 
the virtual $D$ mesons cannot go on-shell and therefore the first term in Eq.~(\ref{prescrip}) which is responsible for the enhancement  of
$X(3872) \to \chi_{c1} \,\pi^0 \,\pi^0$ is absent. However, the decay $\chi_{c1} \, \pi^+ \, \pi^-$ is enhanced relative to 
$\chi_{c0,2} \, \pi^+ \, \pi^-$ decays by factors of $E_\pi/(E_\pi-\Delta_H) \sim 10^{2}$, so we expect
\bea
\left(\frac{{\rm Br}[X(3872) \to \chi_{c 0} \, \pi^+ \, \pi^-]}{{\rm Br}[X(3872) \to \chi_{c 0} \, \pi^0 ]}\right)_{\rm LO}
\approx {\cal O}(10^{-3})\nn \, . 
\eea
(This is the same order of magnitude as the subleading terms in Eq.~(\ref{prescrip}) for the decay $X(3872) \to \chi_{c1}\,\pi^0 \,\pi^0$.)
In the absence of available data on $X(3872) \to \chi_{cJ} \, \pi \, \pi$, we will not attempt a more precise estimate in this paper.

Except for the decay, $X(3872) \to \chi_{c1} \, \pi^0 \, \pi^0$, the decays to final states with $\chi_{cJ}$ plus two pions 
are quite small compared to the two-body decays. These decays are suppressed by factors of $g^2 p^2/\Lambda_\chi^2$ as well as limited phase 
space. In the case of $X(3872) \to \chi_{c1} \, \pi^0 \, \pi^0$, there is an enhancement by a factor of $2 \pi E_\pi/\Gamma_{D^{*0}} \sim 10^4$,
which significantly enhances the decay rate, making it comparable to $\Gamma[X(3872) \to \chi_{c1} \, \pi^0]$.
This enhancement is due to the virtual $D$ meson propagators in Figs.~\ref{Xchipipi} c) and f),
which can go on-shell with $\pi^0 \,\pi^0$ final states.
The potential divergence in this graph is regulated by the tiny width of the $D^{*0}$. 
Ref.~\cite{Dubynskiy:2007tj} observed that the three-body decays to $\chi_{c1} \, \pi\,\pi$ should dominate
over decays to final states with $\chi_{c0}$ or $\chi_{c2}$  when the $X(3872)$ is modeled as
a $^3P_1$ charmonium, however, in this scenario the predictions for the ratios in Eq.~(\ref{lopred}-\ref{nlopred}) 
are quite different. For the hadronic molecule scenario, Ref.~\cite{Dubynskiy:2007tj} applies the QCD multipole 
expansion and finds that the rate for $X(3872) \to \chi_{c0}\,\pi \,\pi$ vanishes, but does not make quantitative 
predictions for the three-body decays involving $\chi_{c1}$ and $\chi_{c2}$.

In summary, we have calculated the decays of the $X(3872)$ to $P$-wave charmonium using a combination
of X-EFT, which treats the $X(3872)$ as a shallow bound state of the nonrelativistic $D$ mesons, and 
HH$\chi$PT. Our methods reproduce established factorization
theorems for decays of the $X(3872)$. HH$\chi$PT can be used to calculate the short-distance coefficients in these factorization theorems.
For two-body decays to $\chi_{cJ} \,\pi^0$ we noted that the $p_\pi^3$ dependence 
that is usually assumed for $P$-wave single pion decays
can be modified by the virtual $D$ meson propagators in the diagrams
shown in Fig.~\ref{Xchipi}. This can affect heavy quark symmetry predictions
for the relative sizes of the partial decay rates. Finally, we find that at LO the partial rates for 
the $X(3872) \to \chi_{c1} \, \pi^0\, \pi^0$ is only slightly smaller than the rate for
$X(3872) \to \chi_{c1} \, \pi^0$, while other decays to final states with $\chi_{cJ}$ and two pions 
are orders of magnitude smaller.

\acknowledgments 

This work was supported in part by the  Director, Office of Science, Office of High Energy
Physics, of the U.S. Department of Energy under Contract No. DE-AC02-05CH11231U.S (SF), Office of Nuclear Physics, of the U.S. Department of Energy under grant numbers DE-FG02-06ER41449 (S.F.), 
DE-FG02-05ER41368 (T.M.), and DE-FG02-05ER41376 (T.M.).
T.M. acknowledges the hospitality of the Particle Theory Group at UCSD
where this work was completed. S.F. acknowledges the hospitality of the Particle Theory Group at LBNL where this work was completed.
We thanks E. Braaten for helpful comments on an earlier version of this manuscript.


\end{document}